\documentclass[prb,twocolumn,showpacs,preprintnumbers,amsmath,amssymb,floatfix]{revtex4}

\usepackage{graphicx}
\usepackage{dcolumn}
\usepackage{bm}
\usepackage{graphicx,rotating}
\usepackage{epsfig}
\usepackage{longtable}

\begin{document}
 
\title{Effect of electron correlations in Pd, Ni, and Co monowires}

\author{
Ma\l{}gorzata Wierzbowska$^{1}$, Anna Delin$^{2}$ and Erio Tosatti$^{1,3,4}$}

\affiliation{$^{1}$INFM DEMOCRITOS National Simulation Center, via Beirut 2--4, 34014 Trieste, Italy\\
$^{2}$Materialvetenskap, Brinellv\"agen 23, KTH, SE-10044 Stockholm, Sweden \\
$^{3}$International School for Advanced Studies (SISSA), via Beirut 2--4, 34014 Trieste, Italy \\
$^{4}$The Abdus Salam International Center for Theoretical Physics (ICTP), Strada Costiera 11, 34100 Trieste, Italy}

\date{\today}
\begin{abstract}
We investigated the effect of mean-field electron correlations on the band electronic
structure of Co, Ni, and Pd ultra-thin monatomic nanowires, at the breaking point,
by means of density-functional calculations in the self-interaction
corrected LDA approach (LDA+SIC) and alternatively by the LDA+$U$ scheme.
We find that adding static electron correlations increases the magnetic moment in Pd
monowires, but has negligible effect on the magnetic moment in Co and Ni.
Furthermore, the number of $d$-dominated conductance channels decreases
somewhat compared to the LDA value, but the number of $s$-dominated
channels is unaffected, and remains equal to one per spin.
\end{abstract}

\maketitle

 \section{introduction}

Nano- and molecular-sized components hold considerable promise
as forming the platform for an entirely new type of hardware technology.
In such a future technology, quantum effects inherent in these
nano-objects  might give rise to new functionalities, possibly
resulting in a totally different design of the basic components
in information technology from what we are used to today.
An important category of systems is constituted by ultra-thin
conductors and nanocontacts. The best known and most investigated
systems in this class are carbon nanotubes and molecular conductors.
Metallic nanowires are also of high interest, but much less studied.
Ultimately thin nanowires, {\sl i.e.} freestanding
atomic chains hanging between tips, constitute a unique metallic structure
where nanosize properties of matter can be tested. Unlike nanotubes, they
are clearly transient objects, although for applications they could be stabilized,
{\sl e.g.} inside other nanostructures. Besides, their transient nature
could itself be put to use.

Metallic nanowires 
are fabricated using a scanning tunneling microscope
(STM) or an atomic force microscope tip. By forming a metallic
contact between the tip and sample of the same metal as the tip,
and then retracting the tip, a thin neck or wire may form by
plastic deformation. Just before it breaks, the nanocontact may
consist of a single atom, and under some circumstances and in some metals
of a short single chain of atoms -- a segment of monatomic nanowire.\cite{ohnishi1998, rodrigues2000}
A second way in which nanowires can be produced is inside a
transmission electron microscope\cite{oshima2003} (TEM).
With the TEM beam, two nearby holes are burned in a thin metal film,
close enough that the bridge between
them narrows down spontaneously to form a nanowire. A third way
to produce nanowires is via
the mechanically controllable break junction technique.\cite{yanson1998}

It was recently predicted that quantum confinement of the
electronic states in metallic atomic chains should lead to a
magnetic ground state, and to a superparamagnetic state at finite
temperatures, in Ru, Rh, Pd, Os
and Pt.\cite{delin_4d,delin_5d,spisak2003,bahn2001}
In the present paper, we build on those results,
also stimulated by some recent intriguing experimental
results indicating unexpectedly low fractional conductance
through Co, Pd and Pt nanocontacts and nanowires observed at room
temperature \cite{ono1999,rodrigues2003,untiedt2004}
(note however that at low temperature, the uncontaminated break junctions
fail to show fractional conductance and even nanowires in the case group
III and IV transition metals.\cite{untiedt2004})
Assuming that, as in the room temperature experiments,
short nanowires could easily form at the nanocontact,
we wish to explore the possible effects of strong correlations
on the electronic structure of the wires, and if and in what
way strong correlations might relate to these recent experimental results.

We will be concerned here only with those correlation effects that can be included
within conventional Fermi liquid theory, where quasiparticle electronic bands
replace the noninteracting electron bands, but can be treated as such in
every other aspect.
Thus, we will not be addressing at all the strongest kind of correlation effects,
 including Coulomb blockades and Kondo phenomena, probably more typical
for quantum dots than for genuine metallic nanocontacts. The effects which
we will address here still amount to a nontrivial renormalization of the
electronic bands, which can however be treated by means of static mean-field
theories such as LDA+$U$ and LDA+SIC.

The electron bands in transition metal nanowires, especially the
$d$ bands, are considerably narrowed by the reduced one-dimensional
coordination. Band narrowing is further enhanced under the strong tensile
stress that brings the nanowires close to the point of breaking.
It is well known that for narrow band systems, standard
density functional theory (DFT) approaches \cite{KS} -- notably
the local density approximation (LDA)
and generalized gradient approximations (GGA) -- 
do not reproduce well the electronic localization within the polarized atomic shells.
For that reason we find it relevant and instructive to
investigate how extensions of DFT, devised specifically to
improve the mean field static description of electron-electron correlations, 
affect the electronic structure of strongly stretched monatomic wires.
In particular, we will make use of the LDA+$U$  approach\cite{ldau_method} and
the self-interaction corrected LDA (LDA+SIC) scheme.\cite{Alessio}

We may anticipate that with these corrections, the band structure
near the Fermi level $E_F$, including bandwidths and splittings
could be strongly altered. In that case also the channel number (number
of propagating bands that cross $E_F$), their transmission, and
in the end ballistic conductance might be sensitive to correlations.
In this work, devoted to strictly ideal infinite monatomic nanowires,
we will address the sensitivity of bands and channels, though clearly
not of transmission, to strong (albeit for the time being statically
described) electron-electron correlations.

 \section{method}
\label{method}

The geometry adopted for the calculation was a two-dimensional lattice of
infinite ideal monatomic chains, with the spacing between chains large enough
to make their interaction negligible. 

We performed all calculations using the SIESTA-1.3 code.\cite{siesta}
The starting point is in all cases a local (spin) density approximation, L(S)DA, 
calculation. On top of that, we treat the effect of strong correlations
by an extension of LDA, namely the LDA+$U$ method \cite{ldau_method} 
and the self-interaction (SI) corrected LDA approach 
(LDA+SIC).\cite{Alessio}
The latter method is an approximated scheme to retain the
exact exchange and it does not add any correlations; LDA+SIC even 
removes a part of correlations, {\em i.e.} self-correlations. 
Thus, the effect of SI correction may differ from the LDA+U result.  

The LDA+U correction potential, $V_{U,m}^{\sigma}$, to the LDA Hamiltonian is diagonal, 
and defined as
\begin{eqnarray}
V_{U,m}^{\sigma} & = & U \sum_{m'} (n_{m'}^{-\sigma}-n_{d}^{-\sigma}) + \nonumber \\
&& (U-J)\sum_{m' \neq m}(n_{m'}^{\sigma}-n_{d}^{\sigma}) +\nonumber \\ 
&& (U-J)(\frac{1}{2}-n_{d}^{\sigma}), 
\label{VU}
\end{eqnarray}
with $n_{d}^{\sigma}$ being the average orbital occupation of the
correlated shell calculated self-consistently from the orbital occupations
$n_{m}^{\sigma}$,  
\begin{equation}
n_{d}^{\sigma} = \frac{1}{2l+1} \sum_{m} n_{m}^{\sigma} = \frac{1}{2l+1} N_{d}, 
\end{equation}
during the LDA+$U$ iteration process. We use here for $N_{d}$ the total actual 
occupation number of the polarization shell (for instance $d$-shell)
rather than fixing $n_{d}^{\sigma}$ at half occupation or using an average LDA occupation.
Details and a test of this implementation are given in Ref.~[\onlinecite{GaMnAsN}].
Our definition of the local occupations $n_{m}^{\sigma}$ also differs from
the Mulliken occupations  $n_{m}^{\sigma,Mulliken}$ 
(Ref.~[\onlinecite{Mull}]) as follows
\begin{eqnarray}
 n_{m}^{\sigma} & = & \sum_{\mu\nu} S_{m \mu}D_{\mu\nu}^{\sigma}S_{\nu m}, 
\label{local} \\
 n_{m}^{\sigma,Mulliken} & = & \sum_{\nu} D_{m \nu}^{\sigma}S_{\nu m},
\label{Mull} 
\end{eqnarray}
involving differently the density matrix $D_{\mu\nu}$ 
and the overlap integrals $S_{m \mu}=\langle  m|\mu \rangle$ and  
$S_{\nu m}=\langle  \nu | m \rangle$ for the basis functions $\mu$ and $\nu$ and
the projector function $m$.  
Further discussion of the local occupation numbers is given in Appendix A.

The $d$-shell exchange parameter $J$ was set to 1~eV in all calculations. 
Varying $J$ in the range $0.5-1.5$~eV was seen to have
only a very minor effect on the studied systems.
Several values for the intra-atomic Coulomb parameter $U$ in the
range $3-9$~eV were tested in order to get a broad insight 
into the electron localization effects. Thus, we use $U$ as
a free parameter in our calculations, although it can in principle 
be calculated self-consistently.
We did calculate
$U$ self-consistently for our nanowires using the 
constrained density functional method described in 
detail in Refs.~[\onlinecite{Matteo}]~and~[\onlinecite{lr1}], in 
addition to treating $U$ as a free parameter.
The results are detailed and critically assessed in Appendix A.
However, an indiscriminate use of self-consistently calculated 
$U$ parameters is not unproblematic.
The LDA+$U$ method has indeed had great success in describing 
systems with rather localized $d$ electrons, 
for example the $3d$ oxide NiO, 
where the LDA+$U$ gives the correct insulating ground state as
opposed to standard LSDA.
On the other hand, when the LDA+$U$ method is applied to
systems which are in effect not strongly correlated, 
for example TiO, it fails, at least when a self-consistently 
calculated $U$ is used in the calculation.\cite{ldau_method}
%

In the present implementation of the LDA+SIC scheme\cite{SIC}, we adopted a modified 
pseudo-SIC approach closely akin but not identical to that in Ref.~[\onlinecite{Alessio}].

The LDA+SIC correction matrix, $V_{SIC,\mu\nu}^{\sigma}$, to the LDA hamiltonian is given by  
\begin{eqnarray}
V_{SIC,\mu\nu}^{\sigma}  & = & \sum_{i} \frac{ \langle \mu | \gamma_{i}^{\sigma} \rangle 
\langle \gamma_{i}^{\sigma} | \nu \rangle }{C_{i}^{\sigma}}, \\
\gamma_{i}^{\sigma}({\bf r}) & = & V_{HXC}^{\sigma}
[\rho_{i}^{\sigma}({\bf r})]\phi_{i}({\bf r}), \\
C_{i}^{\sigma} & = & \langle \phi_{i}| 
V_{HXC}^{\sigma}[\rho_{i}^{\sigma}] |\phi_{i}  \rangle ,
\end{eqnarray}
with $\rho_{i}^{\sigma}({\bf r})$ being the orbital density and the
projectors $\phi_{i}({\bf r})$ being the same basis functions 
$\mu$, $\nu$ used for the calculation of local occupation numbers in the LDA+$U$ method.
Above SI-correction potential differs from the original potential introduced 
by Filippetti and Spaldin  by a factor of 1/2 which we dropped in order to obtain 
the exact result in the one-electron limit ({\em e.g.} the Hydrogen atom). 
The second difference between our approach and the LDA+SIC in Ref.~[\onlinecite{Alessio}] 
is the fact that, we use as a projector the shortest basis function (last zeta-function) 
instead of long-range pseudopotential.
In present work, we restricted the self-interaction correction to the $d$-shell only.

SIESTA treats core electrons by employing the norm-conserving 
pseudopotentials in their fully nonlocal (Kleinman-Bylander) form.\cite{KB} 
We generated the relativistic pseudopotentials for atomic elements using the  
Martins-Troullier\cite{MT} scheme with the nonlinear core corrections.\cite{nlc} 
The electronic configurations and the cut-off radii (in a.u.), in the s/p/d/f order, 
were: $4s^{2}4p^{0}3d^{7}4f^{0}$ and 
2.0/2.0/2.0/2.0 for Co, $4s^{2}4p^{0}3d^{8}4f^{0}$ 
and 1.7/1.88/1.88/1.88 for Ni,
$5s^{1}5p^{0}4d^{9}4f^{0}$ and 2.0/2.2/2.0/2.2 
for Pd in the wire and 2.39/2.39/1.79/1.79 for Pd in the bulk. 

The basis set adopted is a very general and flexible linear 
combination of numerical atomic orbitals (LCAO) \cite{SIESTAbasis} 
by Sankey and Niklewski.\cite{multzeta}
It allows for arbitrary angular momenta, multiple-$\zeta$, polarized and 
off-site orbitals. 
For each atom, we used an automatically generated double zeta basis set  
with polarization functions (which are $p$-type functions in the case 
of transition metals). The second zeta of the $d$-shell was used as 
projector function for the LDA+$U$ and LDA+SIC operators. 
The radii of these projectors were (in a.u.):
1.95 for Ni, 2.073 for Co, and 2.675 for Pd. 
For the self-consistent calculation of the parameter $U$ we tested 
in addition projector radii of 1.8~a.u. for Ni, and 2.2~a.u. for Pd. 

As for the Monkhorst-Pack grid,\cite{mesh} 
for the bulk calculations we used 20$\times$20$\times$20 mesh-points,
and for all wires one {\bf k}-point in the 
plane perpendicular to the wire and 100~{\bf k}-points along 
the wire. This is good enough to represent a single nanowire in 
this case, in case of a very large separation between nearest neighbour
nanowires. Accordingly, the distance between wires in the perpendicular 
plane was set to 10~{\AA}. For the real space grid, we used a uniform mesh 
of quality corresponding to an energy cut-off of 400~Ry in the plane-wave method 
(which, in case of wires with one atom per cell, results to 2187000 mesh-points).

\section{results and discussion}

\subsection{Interatomic distances and magnetization within LSDA}

\begin{figure}
\epsfxsize=7cm
\centerline{
\includegraphics[scale=0.40,angle=0]{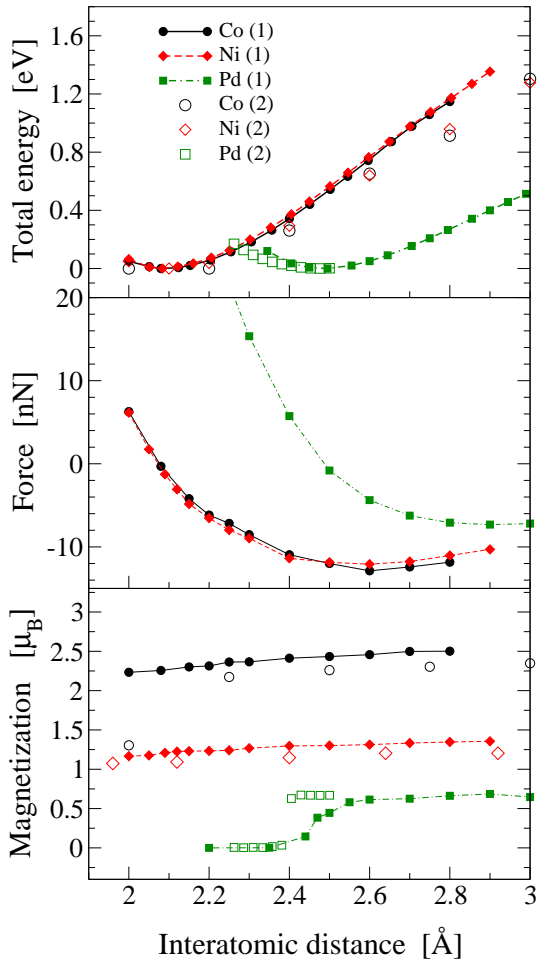}
}
\caption{
\label{figure:etot_force_magn}
Total energy, tension force and magnetization per atom 
for Co, Ni and Pd monowires as a function of 
interatomic distance. 
(1) denotes results obtained with the SIESTA method 
and (2) denotes results obtained with other methods; 
references are given in the text.
The total energy curves for Co, Ni and Pd have been shifted, 
so that the minimum is at zero for all curves.
}
\end{figure}

%
The top panel of Figure~\ref{figure:etot_force_magn}
shows the LSDA total energy as a function of interatomic distance.
Our SIESTA calculations are denoted by closed symbols and 
(1) in the legend.  We see that for all three metals, 
the total energy has a well-defined minimum, which we 
will 
call the equilibrium interatomic distance. 
For both Ni and Co, this distance is around 2.1 {\AA}, 
whereas for Pd, it is significantly larger, around 2.5 {\AA}. 
Decreased coordination from 12 to 2 is of course responsible for
the considerable decrease from bulk interatomic distances, 2.42 , 2.42 and 2.56 {\AA}
respectively for Ni, Co, and Pd.
For comparison, we also show total energy curves obtained
by
other electronic-structure codes (using LDA), 
denoted by open symbols and (2) in the legend. 
For Co and Ni we plotted in particular results obtained by Smogunov {\em et al.} 
in Ref.~[\onlinecite{smogunov}]  by means of plane wave method 
using the PWscf code;\cite{PWscf} for Pd we show results 
of an all-electron FP-LMTO 
(see Ref.~[\onlinecite{FP-LMTO}] for the code) calculation, 
independently performed by us.
As is seen, the agreement between results using the 
SIESTA code and other codes is more than reasonable.

The ideal breaking point $a_b$ of a nanowire is defined as the
atomic distance at which a wire under increasing stress ceases
to remain locally stable and must disappear (if it did not break
earlier at the wire-tip junctions as it happens in practice).
This is an important parameter since in a tip-wire-tip nanocontact
the very last conductance plateau before the wire breaks under increasing
strain will refer to interatomic distances that are larger than 
equilibrium due to stress, but still have $a_b$ as an upper limit.

As defined, the ideal breaking point $a_b$ will coincide with the smallest
interatomic distance associated with instability of some normal mode,
typically with a phonon softening.
While calculating the full phonon spectrum of the wire is
beyond the scope of the present work, we can crudely estimate $a_b$ in a
rather simple way. In a chain held together by pairwise short range forces the
longitudinal vibration branch is unstable when for increasing spacing $a$
the interatomic restoring force is maximally negative,
or equivalently, when the second derivative of the total energy is zero.
Although metal nanowires are not exactly a case of pairwise short range
forces, the crude estimate obtained in that approximation is
sufficient for our purposes.
In the middle panel of  Figure~\ref{figure:etot_force_magn},
we plot the tension force, F, defined as
\begin{equation}
F = p \; A,
\end{equation}
where $p$ is pressure per cell and $A$ is the surface of a plane perpendicular
to the wire and restricted by the walls of the unit cell (in our case this is a hexagon
with an area of 541.265877~\AA$^{2}$, so the pressure of 1~kBar in a cell with 1 atom 
corresponds to the tension force of 5.68~nN).  
The force in the Pd wire is seen to have
a local minimum around 2.9~{\AA}, and the corresponding
values for Ni and Co are  both 2.60~{\AA}. In the following, we will
identify these distances as the ideal breaking points of the wires,
keeping in mind that as explained above, they are really extreme and rough upper limits
of the true breaking points.
We note also that for a Pd monowire the force curve is very flat,
indicating that the interatomic spacing in that system under stress should be
very fluctuating.

\begin{table}
\caption{
\label{table:lattice}
Bulk and nanowire interatomic distances and magnetic moments
calculated with SIESTA in LSDA.
The bulk calculations were performed assuming an ideal
hexagonal structure for Co, and fcc for Ni and Pd.
}
\begin{ruledtabular}
 \begin{tabular}{lccc}
   \\[0.01mm]
  property                                &  Co    & Ni     & Pd      \\[0.2cm]
 \hline
 \hline \\[0.05cm]
\multicolumn{4}{c}{interatomic distance (\AA) }        \\[0.1cm]
 bulk, this work                          &  2.42  &  2.42 &  2.56   \\
 bulk, exp. \cite{Ashcroft}               &  2.51  &  2.46 &  2.67   \\
wire equilibrium distance                 &  2.08  &  2.08  &  2.50   \\
wire breaking distance, $a_b$             &  2.60  &  2.60  &  2.90   \\
\\[0.05cm] \hline \\[0.05cm]
\multicolumn{4}{c}{magnetic moment ($\mu_{B}$) }  \\[0.1cm]
in bulk, this work                        &  1.95  &  0.63  &   0.0    \\
in bulk, exp. \cite{Ashcroft}             &  1.72  &  0.61  &   0.0    \\
at wire equilibrium distance              &  2.23  &  1.21  &   0.45    \\
at wire breaking distance                 &  2.46  &  1.32  &   0.69   \\
free atom configuration    &  $^4 F_{9/2}$  &  $^3 F_{4}$ &  $^1 S_{0}$  \\[0.2cm]
 \end{tabular}
\end{ruledtabular}
\end{table}
%

The bottom panel of Figure~\ref{figure:etot_force_magn}
shows the spin magnetic moment per atom in $\mu_B$ as a
function of the interatomic distance. For Pd, we also
plot the magnetization calculated using the
FP-LMTO code.\cite{delin_4d} The magnetizations obtained for
Ni and Co wires (filled symbols) are compared with
the PWscf calculations given in Ref.~[\onlinecite{smogunov}] (open symbols).
The magnetization results are summarized in
Table~\ref{table:lattice}, where we compare them
with bulk interatomic distances and magnetic moments.

The magnetic moments for Ni and Co calculated with SIESTA are somewhat
higher than these numbers calculated with the
plane wave method, and than experimental bulk magnetizations. 
The onset of magnetism for Ni and Co wires occurs for slightly smaller interatomic
distances in the SIESTA	calculations  
compared to the PWscf calculations. For the Pd wire the magnetism occurs at slightly
larger interatomic distances and appears to be more
gradual than the magnetic moment calculated with the FP-LMTO code.
The source of these slight discrepancies  is  related to 
differences of used pseudopotentials and the difference between the PW and LMTO schemes.
SIESTA uses the normconserving PPs unlike 
the PWscf calculations where ultrasoft PPs were employed. 
We checked also the effect of using the pseudopotentials with and without 
nonlinear core-corrections\cite{nlc} (NLCC) for magnetism. 
For the PPs with NLCC included, magnetism in Pd wire occured at 
the interactomic distance larger of about 0.2~\AA.

All above calculations were performed with unit cells containing only one atom, therefore
the ferrmomagnetic order in wires was assumed. 
%

Finally, we tested the possibility of
antiferromagnetic ordering in Pd, Co and Ni wires.
Antiferromagnetic (AF) exchange could in principle prevail over
direct ferromagnetic (FM) exchange in the extreme
tight binding limit attained for very large $a$.
We performed the AF and FM calculations with two atoms in the unit cell 
(to compare the total energy of two phases). We also calculated 
the chain of seven atoms in the unit cell, flipping the spin in the middle atom.
We found that, collinear antiferromagnetic ordering is 
generally unstable with respect to ferromagnetic ordering, 
at the breaking point distance as well as for shorter interatomic distances,
and in all metals studied in this work.

\subsection{Conductance channels and magnetic moments - 
effect of electron correlations}

A central aim of the present work is to elucidate the effect of
correlations on the number of conductance channels,
foreshadowing in a qualitative way the effect on ballistic conductance.
In the Landauer picture of nanowire conductance, each nanowire
electron band crossing the Fermi level
constitutes a conducting channel between the tips at the two
ends of the nanowire. Each channel can transmit electrons fully
or partially, depending on a transmission coefficient dictated
by the tip-wire junction, and by the character of the electron band.
In transition metals, bands with $s$ character are generally close 
to total transmission.
Owing to their relatively large bandwidth, $s$ electrons are little
disturbed by the potential change
they experience at the tip-wire junction, and are thus essentially
not reflected back, with transmission above 95\%.
In $d$-dominated channels, on the other hand, the smaller bandwidth
of $d$ states leads to higher reflection at the junction, with a
much lower transmission, typically around 20\% (Ref.~[\onlinecite{smogunov}]).

\begin{figure}
\epsfxsize=14cm
\centerline{
\includegraphics[scale=0.33,angle=0]{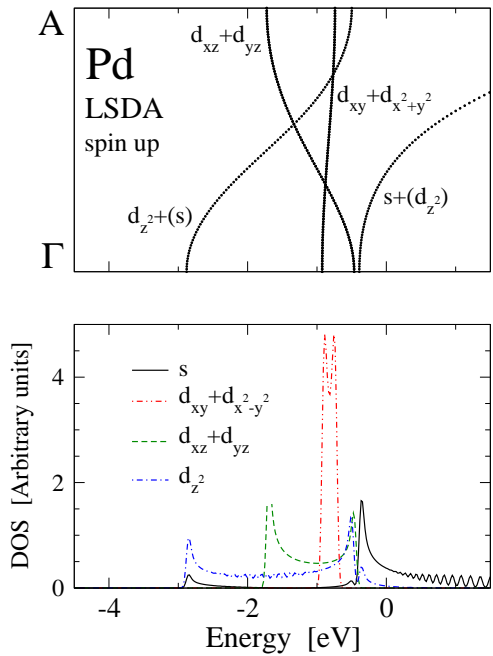}
}
\caption{
\label{figure:symbols}
Nanowire band structure with orbital labels (upper panel) and DOS (lower panel)
for the majority spin of Pd obtained with LSDA at the
breaking point. The Fermi level is at zero.
The symmetry point $A$ in the band-structure plot corresponds to the
point (0,0,$\frac{1}{2}\;\frac{2\pi}{c}$) in the Brillouin zone.
}
\end{figure}

The $d$ bands should of course be narrowest, and correlations strongest,
for a nanowire under maximal strain.
Thus, in order to investigate the effect of electron correlations
on the number of conductance channels of the wires,
we focus here on the nanowires electronic band structures at the breaking points.
These increased interactomic distances are more similar to the conditions 
at which the experiments on breaking junctions have been done. 
Further, we identify the amount of $s$ and $d$ character of the
bands at the vicinity of the Fermi level.
This enables us to draw some qualitative conclusions regarding
the conductance of the wires. In particular, it will allow a discussion of
whether or not a conductance significantly
below the unit conductance quantum $G_0 = 2e^2/h$
is or is not directly suggested by electron correlations, at least within
the present static description.

We start by identifying the character of the bands around the Fermi
level by comparing the majority spin orbital-character resolved
partial density of states (DOS) with the band structure.
In Figure~\ref{figure:symbols}, we show the
band structure and $s$, $d_{z^2}$,
$d_{xy}+d_{x^2-y^2}$, and $d_{xz}+d_{yz}$ partial DOS
for the Pd wire at the breaking point,
obtained from the LSDA calculations.
In the band-structure plot, the character of the individual bands
has been indicated based on information in the DOS plot.
When the band is $s-d$ hybridized but one of the contributions
is much smaller than the other, we show the smaller component within brackets.
The band structures for Co and Ni are generally similar to that of Pd,
and we refrain from repeating the same analysis over and over. For Co and Ni
the character of each band can be obtained by direct comparison with Pd.

\begin{figure*}
\epsfxsize=14cm
\centerline{
\includegraphics[scale=0.40,angle=-90]{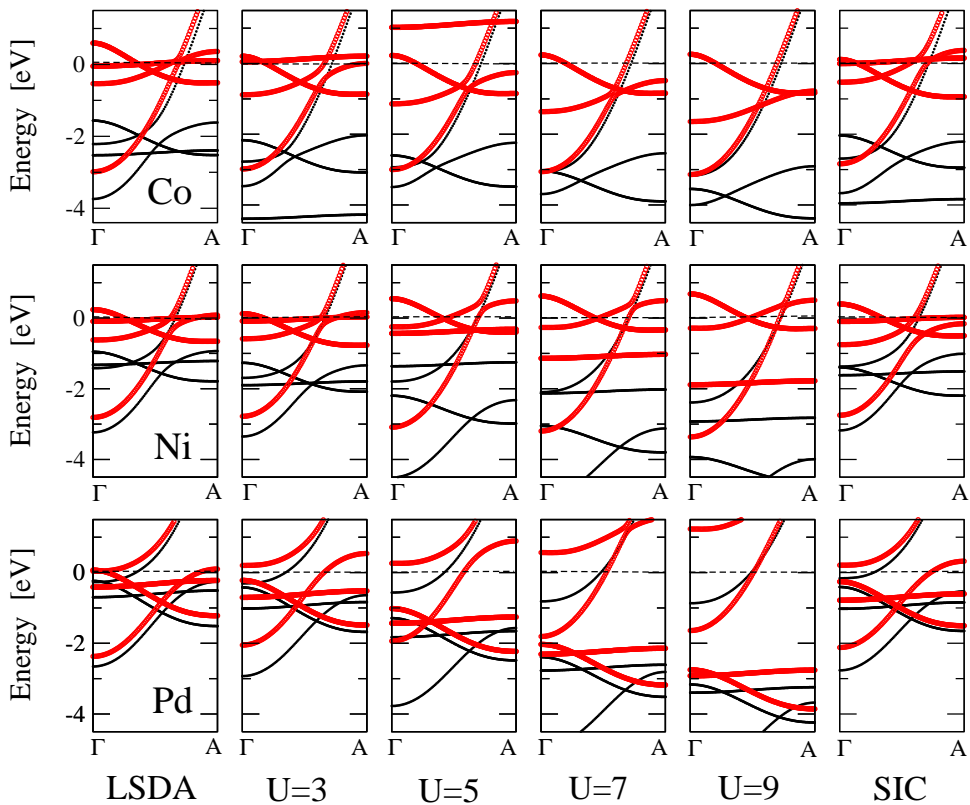}
}
\caption{
\label{figure:band}
Band structures of Co, Ni and Pd nanowires obtained
in LSDA; in LDA+U for $U$ = 3.0, 5.0, 7.0 and 9.0 eV and J=1.0 eV;
and in LDA+SIC ($d$-shell only).
Thin line: spin up electrons; thick line: spin down electrons; dashed line:
Fermi level. All bands are plotted at the breaking point interatomic spacing $a_b$.}
\end{figure*}

\begin{table*}
 \caption{
 \label{table:channels}
Properties of monatomic wires at the breaking point.
$M_T$ is the magnetic moment per atom in $\mu_B$,
$M_s$ the spin-polarization of the $s$ shell in $\mu_B$,
$n_s^{\uparrow + \downarrow}$ the total Mulliken occupation of the
$s$ shell, and $N_c$ the number of spin-resolved conductance channels.}
 \begin{ruledtabular}
 \begin{tabular}{lc ccccc ccccc cccc}
 && & & & && & & & && & & &  \\
  Method   & & \multicolumn{4}{l}{Cobalt} && \multicolumn{4}{l}{Nickel}  &&
              \multicolumn{4}{l}{Palladium}   \\
           & & M$_T$ & M$_s$ & n$_s^{\uparrow + \downarrow}$ &
               (N$_{c}^{\uparrow}$,N$_{c}^{\downarrow}$) &&
            M$_T$ & M$_s$ & n$_s^{\uparrow + \downarrow}$ &
                (N$_{c}^{\uparrow}$,N$_{c}^{\downarrow}$) &&
            M$_T$ & M$_s$ & n$_s^{\uparrow + \downarrow}$ &
                 (N$_{c}^{\uparrow}$,N$_{c}^{\downarrow}$)   \\[0.2cm]
 \hline
 \hline \\[0.05cm]
 LSDA & &   2.46 & $\;$0.03  & 1.40 & (1,6) &
        &   1.32 & 0.03  & 1.24 & (1,6)  &
        &   0.69 & 0.29  & 0.50 & (1,4)  \\
 LDA+U (3 eV) & &   2.44 & -0.01 & 1.43 & (1,6) &
        &   1.32 &  0.03 & 1.24 & (1,6)  &
        &   0.69 &  0.29 & 0.50 & (1,1)  \\
 LDA+U (5 eV)  & &   2.43 & -0.04 & 1.48 & (1,3) &
        &   1.37 &  0.08 & 1.20 & (1,4)  &
        &   0.84 &  0.30 & 0.62 & (1,1)  \\
 LDA+U (7 eV)  & &   2.44 & -0.04 & 1.48 & (1,3) &
        &   1.42 &  0.07 & 1.23 & (1,4)  &
        &   0.96 &  0.22 & 0.80 & (1,1)  \\
 LDA+U (9 eV)  & &   2.45 & -0.03 & 1.47 & (1,3) &
        &   1.44 &  0.05 & 1.26 & (1,4)  &
        &   0.97 &  0.13 & 0.89 & (1,1)  \\
 LDA+SIC & &   2.43 &  $\;$0.01 & 1.42 & (1,6) &
        &   1.28 &  0.02 & 1.23 & (1,5)  &
        &   0.56 &  0.24 & 0.40 & (1,1)  \\[0.2cm]
 \end{tabular}
 \end{ruledtabular}
\end{table*}

Figure~\ref{figure:band} shows the electronic band structures for the Co,
Ni and Pd wires at their respective breaking points.
The left-most column shows first the standard LSDA calculation, the four middle
columns the LDA+$U$ results for a range of $U$ between 3~eV and 9~eV,
and the right-most column the results obtained in LDA+SIC.
The overall effect of added correlations on the electron structures 
is that bands shift in energy and that the dispersion of the $s+d_{z^2}$  
bands change in the vicinity of the Fermi level.
The forms of the individual bands, however, are to a large degree conserved.
We also see that the effects of including self-interaction
corrections on the band structure are roughly similar to the effects of
the LDA+$U$ calculation with $U$ around 3~eV or smaller.

There are quite substantial differences in the way 
LDA+$U$ and LDA+SIC affect the electronic structure.   
The SIC potential is linearly dependent on occupation numbers.
Therefore unoccupied bands do not shift, and nearly empty bands shift very little, 
as for example the minority spin $d_{xy}+d_{x^2-y^2}$ band in Co.
On the contrary, the LDA+$U$ approach moves unoccupied bands upwards 
with a constant shift $(U-J)/2$, due to the last term in Eq.~(\ref{VU}),
(see Section\,\ref{method}).  
As we mentioned in the Section\,\ref{method}, the LDA+SIC and LDA+$U$ results may differ. 
This is due to the fact that LDA+SIC does not add correlations; the method removes "self-exchange"
and "self-correlations". While the LDA+$U$ method is self-interaction free within the considered shell 
and additionally adds correlations within this shell.  

The magnetic moments of Co and Ni are largely unaffected by
increased correlations, whereas the Pd moment increases substantially
with $U$, see Table~\ref{table:channels}.

In general, the properties of transition metal systems are 
strongly sensitive to the relative position of the $s$ and $d$ bands and the
charge transfer between these two bands.
For example, the magnetism in Pd wires is governed by $d \rightarrow s$ charge 
transfer.\cite{delin_4d} 
Therefore, it is of interest to study the effect of correlation on the $d \rightarrow s$ transfer.
Table~\ref{table:channels} shows that
the $d \rightarrow s$ charge transfer decreases with
$U$ for Pd (and thus in principle works toward 
killing off the Pd magnetism by opening up
a gap), but it is much smaller for Ni and roughly constant for Co.

\begin{figure}
\epsfxsize=14cm
\centerline{
\includegraphics[scale=0.60,angle=0]{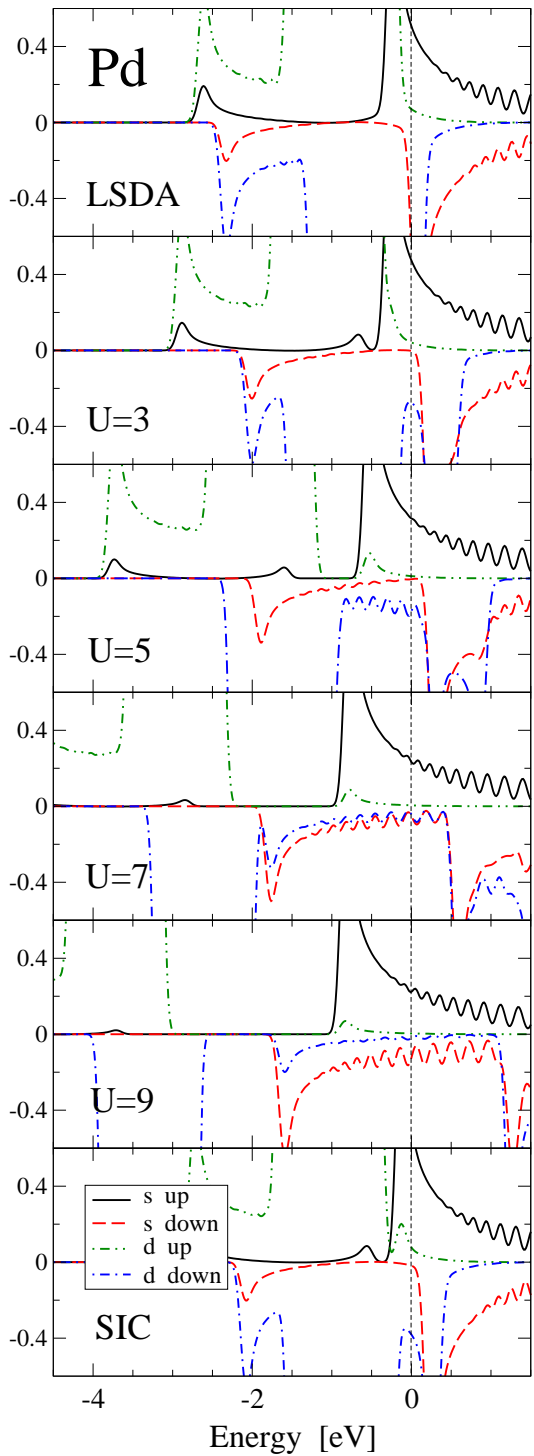}
}
\caption{
\label{figure:sd-ex}
Spin- and orbital-resolved $s$ and $d_{z^{2}}$ DOS for the Pd wire 
at the breaking point, obtained with LSDA (top panel), 
LDA+$U$ (four middle panels) and LDA+SIC (bottom panel). 
The Fermi level is at zero.
}
\end{figure}

%
%

In the Co wire, the largest change of the band structure
when correlations are added is the upward displacement
in energy of the flat nondispersive minority spin $d_{xy}+d_{x^2-y^2}$ band.
The minority channel of this orbital sits right at the
Fermi level in the LSDA and in LDA+SIC calculations, and it is pushed up from
the Fermi level with increasing correlations within the $d$-shell in LDA+$U$.
For $U$=7~eV and above, it is no longer visible in the band-structure plots. 
In contrast to that, the majority spin channel of this orbital is pushed down
in energy. All remaining minority spin orbitals shift slightly down, 
so as to accommodate the charge transfer from the emptied
minority $d_{xy}+d_{x^2-y^2}$ orbital. 
The corresponding majority spin bands are also downward shifted. 
For $U$=3~eV, at the band edge at A, the $d_{z^{2}}+(s)$ band of the minority 
spin channel has moved down slightly and positioned itself just on the Fermi level.
For $U$=5~eV and larger, four bands cross the Fermi level: the minority spin
$d_{xz}+d_{yz}$ (doubly degenerate) and the band $s + (d_{z^{2}})$ for both spins. 
The band $d_{z^{2}}+(s)$ for the minority spin moves down with increasing $U$,
and gradually becomes more $d$-like. 
At the same time, the dispersion and $s$ character of the minority spin 
$s+(d_{z^{2}})$ increases. 
In any case, two $s$ channels are always open irrespective of $U$.
The main difference between the
band structure obtained with the LDA+SIC approach and the
band structure for $U$=3~eV is that at the A edge, 
the minority spin band $s + (d_{z^{2}})$ almost do not move in case of LDA+SIC.
For the other $d$-bands, the effect of SI is very similar to the results obtained with 
the LDA+$U$ method using $U$ of about 2-3~eV. 

Ni has one more electron than Co, and consequently, 
in the LDA+$U$ approach, the $d_{xy}+d_{x^2-y^2}$ (doubly degenerate) 
minority spin bands are now occupied and now shifted downward 
with increasing $U$, in contrast to the situation in Co.
As correlations increase within the $d$-shell, the majority
spin channels with strictly $d$ components are pushed downward.
The partly unfilled minority spin $d_{xz}+d_{yz}$ (doubly degenerate) 
bands of Ni move slightly upward, while the corresponding majority
spin bands move down. This is consistent with a small growth of the total
magnetic moment $M_T$, (see Table~\ref{table:channels}).
For the minority spin bands $s + (d_{z^{2}})$ and $d_{z^{2}}+(s)$ in
Ni, one can see a similar change in dispersion to that
already described for the Co wire.
Also for this metal, two $s$ channels are open at the Fermi level for
all $U$ studied.
%
In the LDA+SIC method, the minority spin band $d_{z^{2}}+(s)$ 
moves a little bit downwards in comparison to the LSDA result, 
unlike the LDA+$U$ method. 

In Pd, the $d_{xy}+d_{x^2-y^2}$ bands move down with increased
correlation within the $d$-shell, just as in Ni.
Already within LSDA, these orbitals are
completely filled, making the magnetic behavior of this
metal very different from the one of Co and Ni.
A more relevant effect of correlation in the $d$-shell
is that the largely empty minority spin channel of the
$s + (d_{z^{2}})$ band is pushed up,
so that it becomes completely unoccupied already for
$U$=3~eV, whereas the corresponding
majority spin channel is pushed down.
In the LDA+SIC band structure, the situation is similar.

Furthermore, the dispersion of the
minority spin channel of the $d_{z^{2}}+(s)$ band
increases so that this band becomes increasingly $s$-like with larger $U$.
For $U$=3~eV or larger, only two bands cross the Fermi level,
one minority spin channel and one majority
spin channel, both of $s+d_{z^{2}}$ character, with the $s$ character
dominating. Thus, also for this metal, the two $s$ channels remain open
in all cases.

In contrast to LDA+$U$,
the LDA+SIC results yield a smaller spin polarization relative
to LSDA within the $d$-shell, and similar magnetization
within the $s$-shell associated with the smaller total
occupation of the $s$-shell.

In Figure~\ref{figure:sd-ex},
we plot the $s$ and the $d$ components of the DOS
for spin channels in the  Pd wire.
In both the LSDA and LDA+SIC calculations, the $s$-component
for both spin-channels are present at the Fermi
level. In the LDA+$U$ scheme with $U$=3~eV,
the minority spin at the Fermi level is dominated by the $d$-component
and the $s$-component almost vanish.
For larger $U$, the $d$-component decreases and the $s$-component grows,
as discussed earlier.  

Due to the aforementioned charge flow between the $s$- and $d$-shells, 
the magnetic moment in the $s$-shell decreases because the $s$-occupation 
in the majority spin is constant and the $s$-occupation in the minority 
spin increases.
While the $d$-occupation in the minority spin decreases and 
the magnetic moment within the $d$-shell increases. 
The above mechanism would keep the total magnetic moment constant, but
the additional contribution comes from the movement of $d$-bands from the Fermi
level (opposite for the majority and minority spins). 
We stress that, the occupation within shells stems from the projected-DOS 
integrated over the whole energy range; therefore one should not conclude 
from Figure~\ref{figure:sd-ex} that a decrease of the projected-DOS at the Fermi 
level is correlated with a decrease of the occupation of the projected shell. 

As for the effect of correlations for the stability, we calculated the equilibrium
and breaking points within LDA+$U$ with $U$=5 eV for all wires.
The equilibrium distances (in \AA) for Co, Ni and Pd are 2.2, 2.11, 2.53 respectively,
and the breaking points are 2.8, 2.63 and 2.96. The total-energy increase from the
equilibrium point to the breaking point calculated within LDA/LDA+$U$ with $U$=5~eV is
(in eV) 0.751/0.666 for Co, 0.769/0.766 for Ni and 0.478/0.331 for Pd.
The decrease of the energetic stability is mostly pronounced in Pd (about 30$\%$) 
and the geometry change is the largest in the Co wire. 
This is somehow consistent with the band structure,
because for Nickel only two bands disappear from the Fermi level when LDA+$U$ with U=5~eV 
is applied while for Pd and Co wires three bands move out from the Fermi level.
   
All above calculations were performed with the unit cells with one atom.
We also calculated chains of seven atoms in the unit cell and with 
the $U$ parameter increasing when going to the middle of a cell. 
This way, we wanted simulate somehow the "tips"; knowing that atoms at tip surface 
are less correlated than atoms in a chain and more correlated than atoms in bulk. 

In summary, for all three metals we find that
both $s$ channels (minority and majority spin) have finite
fractional occupation for all values of $U$ tested and also for the LDA+SIC
calculations, which implies that at least the two conducting $s$ channels are
always present at the Fermi level.

\section{Discussion and conclusions}

In summary, we have found that inclusion of static repulsive electron
correlations within LDA+$U$, while altering noticeably the electronic structure of
maximally stretched monatomic transition metal nanowires, does not affect radically the number
and the nature of conducting channels. In particular, a decrease is found in
the (poorly conducting) $d$ channel number (Table II), whereas the calculated number
of highly conducting $s$ channels remains in all cases one per either spin.
Therefore, one quantum $G_0 = 2e^2/h$ of $s$ conductance will be approximately
preserved after adding correlations at the static level of LDA+$U$.

A  certain reduction is predicted for the $d$ channel conductance. At the same time,
the nature of electronic states and their dispersion near $E_F$ is also changed,
which implies that tip-wire transmission should also be affected. The transmission
changes are not addressed in this work, where tip-wire junctions are not explicitly
included. However, because in  LDA+$U$ the filled states are lowered in energy
and empty states are raised, the width of partly filled bands is somewhat larger,
and that could in turn increase transmission.

Nanowire magnetism (at zero temperature) is poorly affected by correlations in Co and Ni nanowires, 
whereas a clear increase is predicted in Pd. Antiferromagnetism (and the associated
potentially insulating state) never prevails, at least for realistic $U$ parameters and
for interatomic distances below the estimated breaking point.

It is problematic at this stage, where we have not included and treated
at all the wire-tip junctions, to relate the present results to the experimentally
reported ballistic conductance of Co, Ni and Pd nanocontacts. Nevertheless
we can qualitatively remark that the persistence of one $s$ channel per
spin leads to expect in all cases a conductance larger than one in units of $G_0$. 
While this is in agreement with break junction observations, where the ultimate conductance jump
is around 1.3 in Co (Ref.~\onlinecite{untiedt2004}), 
1.4 in Ni (Ref.~\onlinecite{untiedt2004}), and 1.7 in Pd (Ref.~\onlinecite{Csonka}), 
it does leave the fractional conductance steps $G \sim$ 0.5$G_0$
reported in STM experiments at room temperature completely unexplained.
More work, both experimental and theoretical will be needed to clarify this issue.

\begin{acknowledgments}
M.W. is grateful to Stefano de Gironcoli and Javier Junquera for discussion.
We also wish to thank Ruben Weht for checking some of our results with 
the WIEN code \cite{wien97}, as well as D. Ugarte and C. Untiedt for discussions.
Work in SISSA was sponsored by MIUR FIRB RBAU017S8R004,
FIRB RBAU01LX5H, MIUR COFIN2003 and PRIN-COFIN2004, as well as by INFM's
 ``Iniziativa Trasversale Calcolo Parallelo''.
A.D. acknowledges financial support from VR (Swedish Science Foundation).
\end{acknowledgments}

\begin{appendix}
\section{Self-consistent calculation of $U$}
\label{res-3}

We calculated $U$ self-consistently for the wires. 
But also, we made a scan of the
wire properties as a function of $U$, thus, using $U$ as a free parameter.
The main reason for this was that the calculated $U$ 
comes out to be very large.
When calculating $U$ self-consistently,
it is important to notice that the effect of $U$ parameter 
on the band shift and on the other properties,
in any implementation of the LDA+$U$ scheme, depends on the radius of the
projector function, $m$, used to calculate the on-site occupation numbers, 
$n_{m}$,  (see Eq.~(\ref{local})).
And $U$ itself also depends on the projector radius, 
if it is calculated self-consistently. 
For smaller projector radius, calculated $U$ is 
larger and the on-site occupation numbers are smaller. 
Therefore, the combined effect on the correction
to the potential is more/less independent on the projectors. 

The self-consistent $U$ cannot be
directly compared with the Hubbard-$U$, 
because the latter is defined directly by the spectroscopic
properties, {\sl i.e.} the electron affinities and the ionization 
potentials, as follows
\begin{equation}
U = E(N+1)+E(N-1)-2E(N),
\label{Hub}
\end{equation}
where E is the total energy of the system with $N$ electrons in the considered shell. 
The above definition does not depend on the implementation. 
However, using the "exact" spectroscopic data for the $U$ parameter could
give a wrong picture, if such $U$ was combined with the "not exact", 
{\sl i.e.} projector dependent, occupation numbers.
Here we want to note that, for the spectroscopic definition of $U$ we need to use atomic
configuration which involves both $s$- and $d$-shell.\cite{atom-book} Such calculations
of $U$ for atoms were performed for instance by Brandyopadhyay and Sarma\cite{Sarma} 
by means of the Hartree-Fock-Slater method. The phylosophy of standard LDA+$U$ method 
applied in this work, however, is to correlate the electrons in the $d$-shell and not
include the intershell interactions. Satisfying the above condition means that 
we cannot directly compare our $U$ parameters with the atomic Hubbard-$U$, even if we abstract
from the implementation and the combined effect of $U$ and $n_{m}$.  

Focusing on the solid state methods, the Coulomb parameter $U$ can be determined from the 
linear response calculations with respect to a "suitable" perturbation in the 
occupation numbers ("suitable" means small but not too
small, such that the method is sensitive to them). The details of such an
approach are given in Refs.~[\onlinecite{Matteo,lr1}]. 
In this method, the $U$ parameter is equal to
 \begin{equation}
U = (\chi_{0}^{-1}-\chi^{-1})_{II} \nonumber
\end{equation}
where index $II$ denotes the diagonal element at a site $I$, 
which is the atom of our interest,
and the noninteracting and the interacting response functions, 
{\sl i.e.} $\chi_{0}$ and $\chi$, are defined as
\begin{eqnarray}
\chi_{IJ} = \frac{\partial n_I}{\partial \alpha_J}, &\;& 
\chi_{IJ}^{0} = \frac{\partial n_I}{\partial \alpha_J^{KS}}.   \nonumber
\end{eqnarray}
The quantity $n_I$ is the calculated local occupation number 
of the considered polarization shell of the atom $I$.
And $\alpha_I$ is the perturbation of the hamiltonian 
which enters the total energy as follows:
\begin{equation}
E^{KS}[\{ q_I \}] = min_{n({\bf r}),\alpha_I} \left\{ E^{KS}[n({\bf r})] +
\sum_I \alpha_I^{KS} (n_I-q_I) \right\} \nonumber
\end{equation}
with the constrained occupation $q_I$.
The $U$ parameters obtained this way are given in  Table~\ref{table:Uparameters} 
for the Ni, Co and Pd wires and for the separated atoms. 
For Ni and Pd, we performed the calculations at two projectors radii. 

\begin{table}
 \caption{
 \label{table:Uparameters}
Self-consistent $U$ and the local occupations 
$N_{d}$, as well as the Mulliken occupation numbers 
$N_{d}^{Mulliken}$ for the majority spin $\uparrow$ 
and the minority spin $\downarrow$ within the $d$ shell, 
calculated for the atoms in wires and for the separated atoms.} 

 \begin{ruledtabular}
 \begin{tabular}{lccc}
 &  & & \\
  parameters  &     Ni    & Co     & Pd      \\[0.2cm]
 \hline
 \hline \\[0.05cm]
 projector radius (a.u.)  &  1.8\;/\;1.950 &   2.073  &  2.2\;/\;2.675   \\[0.4cm]
\multicolumn{4}{c}{Calculations for wires at the breaking point} \\[0.4cm]
 self-consistent $U$ (eV)  & 13.9\;/\;12.8  &  12.7   &  9.6\;/\;7.6     \\[0.1cm]
 $U\;N_{d}^{\uparrow}$ (eV) & 56.0\;/\;53.5 & 53.0  &  34.8\;/\;31.8  \\[0.1cm]
 $U\;N_{d}^{\downarrow}$ (eV) & 39.1\;/\;37.5 & 25.9  & 30.8\;/\;28.5   \\[0.1cm]
 $N_{d}^{\uparrow}$ &  4.03\;/\;4.18    &  4.17     & 3.62\;/\;4.18     \\[0.1cm]
 $N_{d}^{\downarrow}$ &   2.81\;/\;2.93  &  2.04       & 3.21\;/\;3.75 \\[0.1cm]
 $N_{d}^{\uparrow,Mulliken}$ &  4.91\;/\;4.91   &   4.90  & 4.89\;/\;4.89       \\[0.1cm]
 $N_{d}^{\downarrow,Mulliken}$ & 3.53\;/\;3.54  &  2.50  &  4.39\;/\;4.43  \\[0.1cm]
 $N_{d}^{\uparrow}/N_{d}^{\uparrow,Mulliken}$ ($\%$) & 82\;/\;85  &  85   & 74\;/\;85  \\[0.1cm]
 $N_{d}^{\downarrow}/N_{d}^{\downarrow,Mulliken}$ ($\%$) & 80\;/\;83 & 82   & 73\;/\;85  \\[0.4cm]
 \multicolumn{4}{c}{Calculations for atoms (separation 5~\AA\; in a chain)} \\[0.4cm]
 self-consistent $U$ (eV)    & 16.8\;/\;16.2  &  15.6   &  11.1\;/\;10.7  \\[0.1cm]
 $U\;N_{d}^{\uparrow}$ (eV) & 70.4\;/\;70.3 & 66.8 & 40.4\;/\;44.5   \\[0.1cm]
 $U\;N_{d}^{\downarrow}$ (eV) & 41.5\;/\;41.6 &  29.3 & 37.3\;/\;43.2  \\[0.1cm]
 $N_{d}^{\uparrow}$ &  4.19\;/\;4.34    &  4.28   & 3.64\;/\;4.16   \\[0.1cm]
 $N_{d}^{\downarrow}$ &  2.47\;/\;2.57 &  1.88   &  3.36\;/\;4.04  \\[0.1cm]
 $N_{d}^{\uparrow,Mulliken}$ &   5.00\:/\;5.00  &   5.00  &  4.98\;/\;4.95     \\[0.1cm]
 $N_{d}^{\downarrow,Mulliken}$ &  3.00\;/\;3.00   &  2.19  &  4.60\;/\;4.82 \\[0.1cm]
 $N_{d}^{\uparrow}/N_{d}^{\uparrow,Mulliken}$ ($\%$) & 84\;/\;87  &  86   & 73\;/\;84 \\[0.1cm]
 $N_{d}^{\downarrow}/N_{d}^{\downarrow,Mulliken}$ ($\%$) & 82\;/\;86 & 86  & 73\;/\;84 \\[0.2cm]
 \end{tabular}
 \end{ruledtabular}
\end{table}

In Table~\ref{table:Uparameters}, we collect 
occupations and calculated $U$ parameters.
As one can see, the  value of obtained $U$ parameter decreases with
increasing radius of the projector.
The local occupation numbers within the $d$-shell, $N_{d}$, 
are smaller than the Mulliken ones, $N_{d}^{Mulliken}$.
The Mulliken occupation numbers were obtained by summing over 
contributions from all zeta functions in the basis
set and they almost do not depend on the radius of the 
last (smallest) zeta-function (see Eq.~(\ref{Mull})).
While the local occupations were summed over $d$-orbitals 
within the last zeta functions only, and these numbers 
depend strongly on the projector radius (see Eq.~(\ref{local})). 
The Mulliken occupations sum to the total number of electrons in the unit cell, 
while the local occupations do not need to sum to total number of electrons
within the $d$-shell because they represent only the charge around nuclei and not in the
whole unit cell. Our definitions given by Eq.~(\ref{local}) are
similar to the calculation of a charge closed within the atomic spheres 
in the LMTO approach (Ref.~[\onlinecite{ldau_method}]).   
The local occupations are smaller than the Mulliken ones,
by typically 20$\%$ of $N_{d}^{Mulliken}$. 

In Table~\ref{table:Uparameters}, we give also the product $UN_{d}$ (for each spin) 
calculated at given projector radius.
We assume that, the orbital occupations, $n_{m}$, are proportional
to the averaged occupations, $n_{d}$, and to the total occupations, $N_{d}$, as well.
Since the deviations of orbital occupations from the averaged ones, $n_{m}-n_{d}$, 
are typically one or two orders of magnitude smaller than the occupation numbers 
$n_{m}$ and $n_{d}$ and $N_{d}$, the potential $V_{U}$  (see Eq.~(\ref{VU}))
is much smaller than the numbers $UN_{d}$ presented in Tab.~\ref{table:Uparameters}. 
Nevertheless, the trend for the dependence of $V_{U}$ and of $UN_{d}$ on the projector radius 
is the same, {\sl i.e.} very weak. 
Thus, we may not bother too much what projector radius we chose, 
because the calculated properties will not be much dependent on it, 
as long as it is in the reasonable range from  0.4 to 0.5 of the interatomic distance.

The calculated values of the $U$ parameters seem to be too large. 
But,  we will show that, these numbers give an accurate estimation of the band shift 
calculated with the localized basis set code.
We will use data for NiO (in the rhombohedral structure 
with the antiferromagnetic phase) where we know exactly
the experimental band gap which is 4.3~eV (Ref.~\onlinecite{NiO-exp}). 
We calculated the fundamental gap 
in NiO with LSDA and LDA+$U$ for $U$=2~eV and $U$=8~eV, 
and we obtained 0.22~eV, 0.69~eV, and 3.32~eV respectively, with
the radius of the projector fixed at 2.5 a.u. 
Fitting a parabola to above results, we see that, 
we should use the $U$ parameter of about 9.5~eV to obtain 
the experimental band gap. Such value of the $U$ parameter is
already surprisingly high, not saying about 
values in Tab.~\ref{table:Uparameters} if we 
compare them to other implementations
of the LDA+$U$ scheme. 

We give some reasons for using higher $U$ than typical values used
in the plane wave approach. We rescale linearly the occupations 
for Ni obtained with the projector radius 1.8 a.u. and 1.95 a.u. 
(see Tab.~\ref{table:Uparameters}) to those (spin up and spin down) 
localized occupations which we would obtain using the 
radius of 2.5 a.u.; they would be 4.85 and 3.31 for the majority spin and 
the minority spin respectively.
Now, if we assume that the product $UN_{d}$ 
scales more/less linearly with the projector radius, 
we should take a value of about 48 for $UN_{d}^{\uparrow}$ and of 
about 34 for $UN_{d}^{\downarrow}$ at the radius of 2.5 a.u.  
Dividing these numbers by the expected occupations at that 
projector radius, we will obtain the $U$ parameters of about 9.9~eV and 10.2~eV 
for the majority spin and the minority spin respectively. 
Which gives in average about 10~eV.
We know that the $U$ parameter should be larger for 
the atom than for the bulk. Thus for the wire, a value of 10~eV is quite
good in a comparison to its bulk value of 9.5~eV.
This way, using the experimental data for a band 
gap in NiO and a fit for our numerical data obtained 
with LSDA and LDA+$U$ with $U$=2 and $U$=8~eV, and rescaling 
results for Ni wire to the expected result for a 
larger projector radius, we obtained
two very close values of the $U$ parameter for Ni.
The result of above procedure is the proof that 
one could use the parameters presented in 
Table~\ref{table:Uparameters}
and that obtained this way band shifts in wires would be correct.

On the other hand, the band shift is not the only 
property one would like to get from the LDA+$U$ scheme. 
For instance, the correct orbital ordering in NiO could 
be obtained with a smaller value of $U$, about 6~eV. 
Therefore, for the sake of the balance between 
different properties which we want to describe, we take a bit 
smaller values of $U$ than these which give the expected 
band shifts in bulk. 

At the end we conclude that, for the number of channels 
crossing the Fermi level, it is not very important how large $U$ we take, 
because for $U$ about 3$\sim$4~eV the bands which are supposed to 
move already have done it.

\end{appendix}

\end{document}